\documentclass[useAMS,usenatbib,usegraphicx,longnamesfirst]{mn2e}

\usepackage{amsmath2000,times}

\title{Degeneracies and scaling relations in general power-law models for
gravitational lenses}

\author[O.~Wucknitz]
{Olaf~Wucknitz\thanks{e-mail: {\tt
owucknitz@hs.uni-hamburg.de}}
 \\
  Universit\"at Hamburg, Hamburger Sternwarte, Gojenbergsweg 112,
  21029 Hamburg, 
  Germany \\
  University of Manchester, Jodrell Bank Observatory, Macclesfield,
  Cheshire SK11 9DL, UK}

\date{21 November 2001}

\renewcommand{\epsilon}{\varepsilon}

\newcommand{\vc}[1]{\text{\boldmath$#1$}}

\newcommand{\psig}{\psi\supp{(g)}}

\newcommand{\psigam}{\psi^{(\gamma)}}

\newcommand{\rtext}[1]{\quad\text{#1}}

\newcommand{\sub}[1]{_{\mathrm{#1}}}

\newcommand{\supp}[1]{^{\mathrm{#1}}}

\newcommand{\zs}{z\sub s}

\newcommand{\sprod}{\cdot}

\newcommand{\matr}[1]{\begin{pmatrix}#1\end{pmatrix}}

\DeclareSymbolFont{CMSSB}{U}{cmss}{bx}{n}
\DeclareMathSymbol{\matGamma}{0}{CMSSB}{"0}

\newcommand{\matg}[1]{{\let\Gamma=\matGamma#1}}

\newcommand{\transp}[1]{{#1}^{\text{t}}}

\newcommand{\diff}{{\rmn{d}}}

\newcommand{\pam}[2]{\begin{smallmatrix} +#1 \\ -#2\end{smallmatrix}}

\newcommand{\twovec}[2]{\left(\,#1\,,\,#2\,\right)}

\newcommand{\twovectext}[2]{(#1,#2)}

\newcommand{\h}{h}

\newcommand{\hsph}{h\sub{sph}}

\newcommand{\hnogam}{h_0}

\newcommand{\kmsmpc}{\rmn{km\,s^{-1}\,Mpc^{-1}}}

\shortcites{schechter97,courbin97,impey98,crane91,burud98,bade97,koopmans99i,jackson98,chartas00,chartas01}

\begin{document}

\maketitle

\begin{abstract}
The time delay in gravitational lenses can be used to derive the
Hubble constant in a relatively simple way \citep{refsdal64}. The
results of this method 
are less dependent on astrophysical assumptions than in many other
methods. For systems with accurately measured positions and time
delays, the most important uncertainty is related to the mass
model used. Simple parametric models like isothermal ellipsoidal mass
distributions seem to provide consistent results with a reasonably
small scatter when applied to several lens systems
\citep{koopmans99ii}. We discuss a family of models with a separable
radial power-law and an arbitrary angular dependence for the
potential $\psi=r^\beta\,F(\theta)$. Isothermal potentials are a special case of these models
with $\beta=1$. An additional external
shear is used to take into account perturbations from other
galaxies. 
Using a simple linear 
formalism for quadruple lenses, we can derive $H_0$ as a function of
the observables and 
the shear. If the latter is fixed, the result depends on the assumed
power-law exponent according to $H_0\propto(2-\beta)/\beta$. The
effect of external shear is quantified by introducing a `critical
shear' $\gamma\sub c$ as a measure for the amount of shear that
changes the result significantly.
The analysis shows, that in the general
case $H_0$ and $\gamma\sub c$ do not depend on the position of the
lens galaxy. 
Spherical lens models with images close to the Einstein radius with
fitted external shear differ by a factor of $\beta/2$ from
shearless models, leading to $H_0\propto 2-\beta $ in this case.
We discuss these
results and compare with numerical models for a number of real
lens systems. 
\end{abstract}


\section{Introduction}

Gravitational lenses offer a unique possibility to determine
cosmological distances and hence the Hubble constant \citep{refsdal64}. The
method avoids using any form of distance ladder and is almost
independent of a deeper understanding of astrophysical processes. This
has the advantage that possible statistical and systematic
uncertainties can be controlled much better than with other methods.
Besides the measurement errors of redshifts, positions and time
delays, the most important source of uncertainties in using
gravitational lenses to determine $H_0$ is the mass model of the
lens. Constraints provided by observational data are never sufficient
to fix the mass distribution uniquely \citep{saha97,williams00}. Simple
parametric models, based on the knowledge about typical galaxies in the
local universe or desired mathematical properties, are normally used to
overcome this difficulty.
This approach, while leading to consistent results, has the
disadvantage of hiding the underlying uncertainties and making it
difficult to quantify them.

In this paper, we consider a family of mass models with separable radial
and angular dependence of the potential. External perturbations are
included as an external shear. In
this way, the different parts can be studied independently. For the radial
dependence, we choose a simple power-law and are especially interested
in the influence the radial slope has on the results.
Several authors \citep{wambsganss94,witt95,wucknitz99} have found
approximate scaling relations of $H_0\propto2-\beta$ for spherical models with
external shear. The same dependence was found before by
\citet{chang76} for doubles in spherical
models without shear \citep[see also][]{refsdal94}. 
The work of
\citet{williams00} showed, that the radial slope also has important
effects in non-parametric models. Due to the very general nature
of these models, no exact scaling law could be found for them.

We follow an intermediate approach by using an arbitrary function only
for the angular part. The models include
elliptical mass distributions and potentials as well as other
shapes. It has been observed in several lens systems, that the
external shear required to fit the data with simple elliptical models
is much higher than expected. This might indicate a kind of asymmetry
of the galaxy that cannot be accounted for by elliptical
models. The general angular part of the potential we use can
describe this `internal shear' more accurately than simple parametric
approaches. 

We extend and generalize the work of \citet{witt00} by including the
three independent time delays in quadruple systems as constraints for
the models.
We do not
use magnification ratios for several reasons. One is the
observational problem of reliably measuring correct flux
ratios. Fluxes are influenced by microlensing and extinction. These
effects can be very strong in the optical and in some cases they are
still significant for radio wavelengths. The other reasons are related
to our formalism and will be discussed later. 

With this approach, we can derive explicit solutions for the Hubble
constant as a 
function of the observables, the power-law exponent $\beta$, and the external
shear.
We use the results to find a simple but rigorous scaling law describing the
dependence of $H_0\Delta t$ on $\beta$ in lenses with quadruple images.
This scaling law of $H_0\Delta t\propto(2-\beta)/\beta$ is also valid for more
special parametric power-law 
models within the allowed range of $\beta$ and is therefore inherent
to these models and not an artefact of the general angular
dependence we discuss.

The qualitative effect of this scaling law is easily understood when
comparing it with the $2-\beta$ scaling discussed before and with the
mass-sheet degeneracy, which leads to $H_0\Delta t\propto 1-\kappa$. In all
cases, a shallower density profile (larger $\beta$ or $\kappa$) leads
to smaller values of $H_0\Delta t$.
The flat part of the density distribution ($\kappa$ in the
mass-sheet degeneracy) amplifies the deflection angles but leaves the
time delays unaffected. To fit the given geometry, the total deflection angles
have to be constant, therefore the time delays ($H_0\,\Delta t$) decrease.

We also discuss the effect of the external shear on the time delays
and the measured Hubble constant. If the shear is changed, the
internal asymmetries of the mass distribution have to be adjusted to
compensate for the changes. The total effect of these two contributions
will be described by the new concept of a `critical shear'
$\vc\gamma\sub c$. The measured Hubble constant is a linear function
of the external shear and becomes zero for $\vc\gamma=\vc\gamma\sub
c$. We will present a simple interpretation of the critical shear in
terms of the image geometry.

The main goal of this work is to investigate the uncertainties in
measurements of $H_0$ or cosmological parameters from time delays in
gravitational lenses. The combination of distances that can be
determined not only depends on the redshifts and $H_0$ but also on
the other cosmological parameters ($\Omega$ and $\lambda$ in standard
cosmology) and the smoothness of the mean density distribution
(`lumpiness' parameter $\alpha$). The results will also help in minimizing
possible errors, either by selecting lenses with the least
uncertainties or by using constraints of the model
parameters which are lensing-independent. 

Finally, we apply the formalism to several known lens systems, with and without
measured time delays, to compare our analytical results with those from
numerical model fitting using parametric models.
For 2237+0305, we compare with own numerical
models and find a very good agreement even though the time delays are
not used as constraints in this case.

Our results can be used directly to determine $H_0$ from
time delays without explicit modelling but should not be used as a
substitute for it. They are rather meant to provide an
explanation for the degeneracies and scaling relations that have been
found for several families of lens models.
Nevertheless, we show that the direct application to real systems
is possible.

In the appendix, we discuss possible Einstein rings in our family of
models. Interestingly, infinitely small sources can be mapped as
elliptical rings for arbitrary values of the external shear. The
ellipticity of these rings is directly determined by the shear.

\section{The lens model}

For the lens equation and deflecting potential $\psi$, we use the
following notation:
\begin{align}
\vc\zs &= \vc z - \vc\alpha (\vc z)
\label{eq:lens} \\
\vc\alpha(\vc z) &= \vc\nabla \psi(\vc z) \\
\vc z &= r\,\twovec{\cos\theta}{\sin\theta}
\end{align}
We use a power-law approach with a general
separable angular 
part for the main lensing galaxy. 
\begin{equation}
\psig (\vc z) = r^\beta \,F(\theta)
\end{equation}
This family of models includes both
elliptical mass distributions and 
elliptical potentials with arbitrary radial mass index $\beta$.
In the following, we assume all image positions as known. For the
chosen model, this implies also the knowledge of the position of the
galaxy centre itself, because coordinates are relative to this centre.
Later we will see, however, that some of the equations are translation
invariant, leaving the results unchanged when shifting the galaxy.
The limiting cases of $\beta\to0$ and $\beta=2$ correspond to a point
mass and a density constant in the radial direction,
respectively. Between these two, the isothermal models are another
special case with $\beta=1$.
The normalized surface mass density of the general model is
\begin{align}
\sigma (\vc z) &= \frac12 \vc\nabla^2 \psi \\
&= \frac{r^{\beta-2}}{2}\Bigl(\beta^2\,F(\theta)+F''(\theta)\Bigr) \rtext{.}
\label{eq:sigma}
\end{align}
Here and in the rest of this paper, primes indicate derivatives with
respect to $\theta$.
A very simple relation holds for the radial derivative of the
potential which we will need in the time delay equations later:
\begin{equation}
\vc z\sprod\vc\nabla\psig = \beta \, \psig \label{eq:rad deriv}
\end{equation}

To account for nearby field galaxies or the contribution of a cluster,
we include an external shear 
plus an additional constant mass density or convergence $\kappa$ into
our models. This $\kappa$ can not be
determined from observations as a consequence of the so called
mass-sheet degeneracy, first discussed by \citet{falco85} and
\citet{gor88}. In the following, we therefore always use a fixed $\kappa$.
We parametrize the shear as follows:
\begin{align}
\psigam(\vc z) &=  \frac12 \transp{\vc z} \matg\Gamma \vc z \\
\intertext{The shear matrix $\matg\Gamma$ includes both the convergence
  and the shear itself.}
\matg\Gamma &= \matr{\kappa-\gamma_x & -\gamma_y \\ -\gamma_y &
  \kappa+\gamma_x} \\
\label{eq:shear dir}
\vc\gamma &= \gamma \,\twovec{\cos2\theta_\gamma}{\sin2\theta_\gamma}
\end{align}
With this definition, $\theta_\gamma$ points in the direction of the
external perturbing mass or equivalently in the opposite
direction.
Note that this part of the potential is a special case of
the power-law with $\beta=2$.

\section{Time delays}

The light travel time $t$ for a certain image at $\vc z$ can be
calculated for arbitrary lens models by using the equation
\begin{gather}
t = \frac{1}{c} D\sub{eff}
\left( \frac12 |\vc\alpha(\vc z)|^2-\psi(\vc z)
\right) + \text{const} \rtext{with} \\
D\sub{eff} = \frac{D\sub d\,D\sub s}{D\sub{ds}} \,
(1+z\sub{d}) \rtext{.}
\end{gather}
Here $D\sub d$ and $D\sub s$ are angular size distances of the
deflector and the source from the observer, while $D\sub{ds}$ is the
distance of the source measured from the deflector.
We separate this equation as usual into one part depending only on
cosmology (without $H_0$), the contribution from the
Hubble constant, and a lens dependent part. This is done
by using reduced distances $d_x$ (with $D_x = d_x \, c/H_0$) that
only depend on 
the cosmological parameters (and the redshifts) but not on $H_0$ itself.
We now use a scaled version of the Hubble constant,
\begin{equation}
h =\frac{H_0}{d\sub{eff}} \rtext{,}
\end{equation}
which
includes the cosmological factors and will later be determined by the
lensing effect. This $h$ is directly proportional to the Hubble
constant itself when the cosmology is fixed.
For two images $i$ and $j$, we get a time delay of $\Delta t_{ij}$ with
\begin{align}
\h\,\Delta t_{ij} &= h\,(t_i - t_j) \\
&=  \frac12 \left(|\vc\alpha_i|^2-|\vc\alpha_j|^2\right)-
(\psi_i-\psi_j) \rtext{.}
\end{align}
By using the lens equation \eqref{eq:lens}, we can
transform this expression into a linear functional of the deflecting
potential and its derivatives. Mixed terms like $\vc
z_i\sprod\vc\alpha_j$ for $i\ne j$ can be eliminated, so that the
resulting time delay 
can again be written as the difference of the light travel times $t_i$
themselves.
\begin{equation}
\h\,t_i = 
 -\frac12 r_i^2
+ \vc z_i\sprod\vc\nabla\psi_i  -\psi_i -C
\end{equation}
Although it is not immediately noticeable, this equation still is
invariant under translations of coordinates.
To prove this, one has to apply the lens equation.

For the special potential we discuss here, the
relation \eqref{eq:rad deriv}, which also holds for $\psigam$
with $\beta=2$, can be used to eliminate the
derivatives of the potential and obtain a simple
expression only depending on both parts of the potential at the
image position: 
\begin{equation}
\label{eq:time del 2}
\h\, t_i = -\frac12 r_i^2 - (1-\beta) \psig_i + \psigam_i - C
\end{equation}
We retain using the notion of light travel times for each image instead
of time delays {\em between} the images to keep the equations simple. The
$t_i$ are defined except for an overall additive constant
that gets absorbed into the constant $C$.

The last equation was already presented in \citet{witt00} for the
special cases of $\beta=1$ including external shear and for the general
shearless power-law model.

\section{Counting constraints and parameters}

Before solving the equations, we want to discuss how many parameters
of the model can at most be determined from observations of lens
systems with $n$ images, especially $n=4$.
See Table~\ref{tab:parameters} and \ref{tab:constraints} for a list of
parameters and constraining equations.
As the time
delays do not change when adding the 
same constant to all $t_i$, we have to include the constant $C$ as
parameter and thus have $n$ constraints with one parameter for
this. We might as well have used only the (uniquely defined) $n-1$
time delays without $C$. Both possibilities result in $n-1$ more
constraints than parameters just for the time delays.

\begin{table}
\caption{Parameters for the chosen family of models for a system with
  $n$ images. The $F'_i$ are needed in the lens equations.} 
\label{tab:parameters}
\begin{center}\begin{tabular}{clc}
parameters    &                                     & number \\ \hline
$h$           & scaled Hubble constant              & 1\\
$\vc\gamma$   & external shear                      & 2 \\
$\beta$       & power-law exponent                  & 1 \\
$F_i$         & angular part $F(\theta_i)$          & $n$\\
$F'_i$        & $\diff F/\diff\theta$ at $\theta_i$ & $n$ \\
$\vc z\sub s$ & source position                     & 2 \\
$C$           & constant in light travel times      & 1 \\ \hline
total         & without fluxes                      & $2n+7$
\end{tabular}\end{center}
\end{table}%
\begin{table}
\caption{Constraints from
  observations of image positions and time delays for a system with
  $n$ images.} 
\label{tab:constraints}
\begin{center}\begin{tabular}{clc}
constraints&                & number \\ \hline
 $\vc z_i$ & image positions & $2n$ \\
 $t_i$     & light travel times    & $n$ \\ \hline
 total     & without fluxes &$3n$
\end{tabular}\end{center}
\end{table}%

Even for systems with 4 images of one source, it will be impossible
to determine all parameters. We therefore decide to fix $\beta$ in the
following calculations so that the results can be used to study the
dependence of $\h$ on the radial mass slope. We will see that, with
fixed $\beta$, all equations stay linear.

Another critical
parameter is the external shear, which seems to be higher than expected
in many detailed lens models. Fixing $\gamma$ for the moment, we will
be able to 
investigate the influence of any uncertainties in the external shear.
In the special case of $\beta=1$, it will even be possible to
determine $\vc\gamma$ from the constraints, because a number of other
parameters do not contribute then.

We do not include fluxes or magnifications, because
they would provide only $n-1$ constraints (the independent flux
ratios) and at the same time 
add $n$ more parameters (the second derivatives of the angular
part of the potential at the image positions $F''_i$).
The effect, that models get less constrained, when more observations are
included in the analysis, is of course unknown in normal parametric
models, where the number of parameters is fixed. Our models have an
infinite number of parameters, of which only a finite subset is needed
to compare with observations. The number of relevant parameters
can change when we include more constraints.

Another reason
(besides the difficulties in determining fluxes already discussed)
is that, in contrast to 
deflection angles and time delays, magnification
ratios are non-linear functionals of the lensing potential, complicating
the equations considerably.

\section{Lens equations}

To exploit the information given by the image positions, we have
to insert the power-law potential with shear into the lens equation
\eqref{eq:lens}. 
The derivative of the galaxy part of the potential can be
obtained most simply by rotating its polar form to
Cartesian coordinates using the transformation matrix
\begin{equation}
\matr{\upartial_x \\ \upartial_y} 
= \matr{\cos\theta & -\sin\theta/r \\ \sin\theta & \cos\theta/r}
\matr{\upartial_r \\ 
  \upartial_\theta} \rtext{.}
\end{equation}
Written in a form to take into account the role of
$\vc\gamma$, $F_i$ and $F'_i$ as unknowns of the equations, this leads
to the following equation.
\begin{equation}
\vc\zs = (1-\kappa)\,\vc z_i - r_i^{\beta-2} \matr{x_i & -y_i \\ y_i & x_i}
\matr{\beta F_i \\ F'_i} + \matr{x_i & y_i \\ -y_i & x_i}
\matr{\gamma_x \\ \gamma_y}
\end{equation}
It is easily seen, that this set of $2n$ equations can be used to
determine $F_i$ and $F'_i$, assuming $\vc\gamma$ as
known:
\begin{align}
\label{eq:F}
\beta F_i &= r_i^{-\beta} \Bigl( (1-\kappa)\,r_i^2 -x_i x\sub s - y_i y\sub s +
\gamma_x\,(x_i^2 -y_i^2) +2\,\gamma_y\,x_iy_i\Bigr) \\
\label{eq:F'}
F'_i &= r_i^{-\beta} \Bigl( y_i x\sub s - x_i y\sub s -
2\,\gamma_x\,x_iy_i +\gamma_y\,(x_i^2 -y_i^2)  \Bigr)
\end{align}

\section{The general set of linear equations}

We now use \eqref{eq:F} to express the galaxy
potential in the time delay equations \eqref{eq:time del 2}. We
decide to use the light travel times themselves rather than the
$H_0$-independent time delay ratios as constraints. In this way we can
keep the equations 
linear and the analysis much simpler. As a
result, the following set of equations is obtained for $i=1\dots n$:
\begin{equation}
\begin{split}
  (1-\kappa)\,r_i^2
  \;&+\; (x_i^2-y_i^2)\,\gamma_x \;+\;
2x_iy_i\,\gamma_y \\
\;&+\;\frac{2\beta}{2-\beta}\,(t_i\,\h +C)
 \;=\; 2\frac{1-\beta}{2-\beta}\,\vc
z_i\sprod\vc\zs
\end{split}
\label{eq:eqset comp}
\end{equation}
The most interesting fact, besides the linearity, is the simple way
the mass index $\beta$ appears in the equations. By combining
information from the time delay and lens equations the way we did, it
was possible to remove the terms with $\beta$-dependent 
exponents. Now $\beta$ only contributes in the scaling factors of the
unknown parameters. Having solved the system for one value of $\beta$, we
immediately find the general solution by scaling $\h$, $\vc\zs$ and
$C$ with the appropriate factors.

\section{The isothermal model}
\label{sec:iso}

In the case $\beta=1$, the equations \eqref{eq:eqset comp} obviously
degenerate with respect to the source position $\vc\zs$. It is then
impossible to constrain the latter, but 
the same information can now be used to determine the external shear.
In this case, the inclusion of the lens equations to determine
the parameters $F_i$ of the galaxy potential was not really
necessary, because $\psig$ does not contribute to the time delay
equations \eqref{eq:time del 2}. We can now directly invert the latter
or \eqref{eq:eqset comp} to obtain solutions for
$\h\sub{iso}$ and $\vc\gamma\sub{iso}$.

The transition from exactly isothermal to nearly isothermal models deserves
some attention. Even for models with $\beta$ differing only slightly
from unity, $\vc\gamma$ is a free parameter, while it is fixed by the
observational data for $\beta=1$. An incorrect external shear 
for almost isothermal models is usually compensated for by diverging source
positions and $F_i$ and $F'_i$, leading to unrealistic
models. It therefore seems appropriate to use the correct
isothermal shear even for models that do not exactly obey
$\beta=1$. However, one has to take into account possible measurement
uncertainties that introduce errors into
$\vc\gamma\sub{iso}$. Especially the time delays, which all contribute
to the solution, can introduce significant uncertainties.

\section{Solution for the general model}

In the general case with $\beta\ne1$, \eqref{eq:eqset comp} can be
solved directly to determine $h$ for a given shear $\vc\gamma$.
Even without writing the explicit solution, we see, that
\begin{equation}
h \propto \frac{2-\beta}{\beta} (1+g_x\,\gamma_x+g_y\,\gamma_y)
\label{eq:H' general new}
\end{equation}
for some constants $g_x$ and $g_y$.
The Hubble constant scales with $(2-\beta)/\beta$ and is a linear (but not
proportional) function of the shear.
Notably, the isothermal model plays no special role in this equation,
despite the fact that, strictly speaking, $\vc\gamma$ can not be chosen
freely in this case. As the isothermal shear is usually only weakly
constrained due to the limited measurement accuracy, we may, however,
use different values of $\vc\gamma$ even for $\beta=1$. This is
always true when the constraints we used here are not all
available. Considering this, one might well apply equation
\eqref{eq:H' general new} 
regardless of the value of $\beta$.

\section{The `Critical shear'}

As the Hubble constant is linear in $\vc\gamma$, there has to be a
one-dimensional family of values of the shear with vanishing $h$.
The shear with the smallest absolute value from this family
will now be called the `critical shear' $\vc\gamma\sub c$.
If we denote the shearless value with $\hnogam$, we can write the
effect of external shear as
\begin{equation}
\frac{\h}{\hnogam} = 1-\frac{\vc\gamma \sprod \vc\gamma\sub
  c}{\gamma\sub c^2} \rtext{.}
\label{eq:h crit shear}
\end{equation}
If $\vc\gamma$ and $\vc\gamma\sub
c$ point in the same direction, this scaling factor can be written as
$1-\gamma/\gamma\sub c$ which is analogous to the scaling of
$1-\sigma/\sigma\sub c$ for an additional mass sheet $\sigma$ with critical
density $\sigma\sub c$.
The critical shear does not dependent on $\beta$ or the time delays
$t_i$ and can be calculated from the image positions alone.

If an upper bound of $\gamma\sub{max}$ can be assumed, this
translates to a range of
\begin{equation}
\h\sub{\bigl(\begin{smallmatrix} \rmn{max}\\\rmn{min}\end{smallmatrix}\bigr)} = \left(1\pm\frac{\gamma\sub{max}}{\gamma\sub c}
\right) \hnogam \rtext{.}
\end{equation}
The critical shear is thus a measure for the amount of external shear that
can change 
$\h$ significantly. The larger it is, the smaller the influence of
uncertainties in the shear on the determined Hubble constant. Shear in
the direction of $\vc\gamma\sub c$ contributes maximally, in a
perpendicular direction the influence vanishes. With the `direction
of shear', we mean in this context the orientation $2\theta_\gamma$ of the
vector $\vc\gamma$. This is {\em not} the direction of $\theta_\gamma$
towards the perturbing mass, cf.\ eq.\ \eqref{eq:shear dir}. External
field galaxies located 
perpendicular to the direction of maximal effect change $\h$ by the
same amount but with opposite sign.
There are, however, four directions where external masses do not contribute.
See section \ref{sec:special cases} for actual numbers of the critical
shear in real observed lens systems.

\subsection*{Geometrical interpretation}

\begin{figure}
\begin{center}\includegraphics{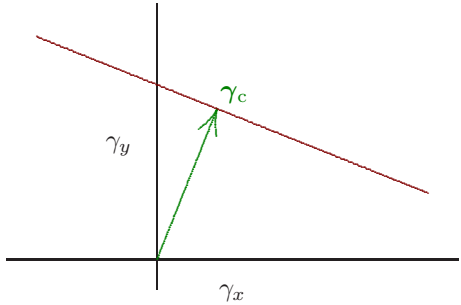}\end{center}
\caption{Example critical shear $\vc\gamma\sub c$ (arrow)
  and values of $\vc\gamma$ for which $h/\hnogam=0$ (straight line).
  Note that the direction of $\vc\gamma$ is $2\theta_\gamma$ and \emph{not}
  the direction to the perturbing mass $\theta_\gamma$.}
\label{fig:gammac}
\end{figure}

\begin{figure}
\begin{center}\includegraphics[width=8.3cm]{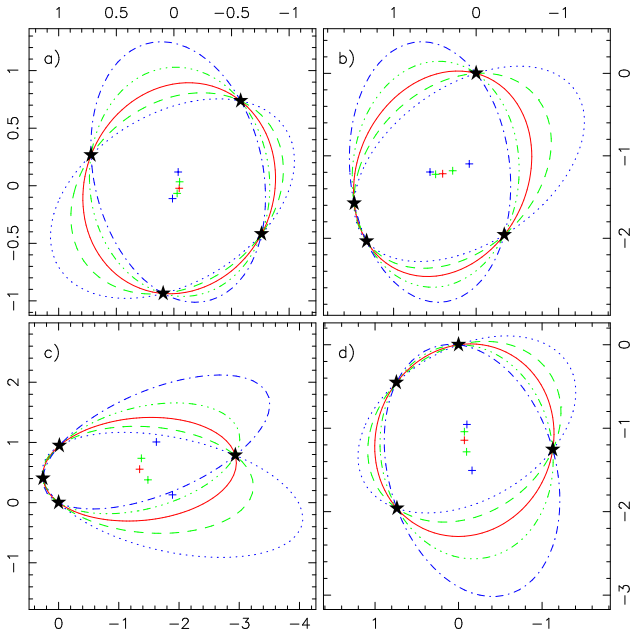}\end{center}
\caption{Ellipses passing through all images (stars) of four lens
  systems. The `roundest' one, which corresponds to the critical
  shear, is shown as solid line. The centres are shown as crosses. For
  each ellipse, a shear for which $h/\hnogam=0$ can be calculated by
  using equation \eqref{eq:gamma a2b2}. a) Q2237+0305, b) PG~1115+080,
  c) RX~J0911.4+0551, d) B1608+656} 
\label{fig:crit shear}
\end{figure}

For constant light travel times or $h=0$, equation \eqref{eq:eqset comp}
describes an ellipse
whose axes $a$ and $b$ are related to the external shear by
\begin{equation}
\frac{\gamma}{1-\kappa} = \frac{a^2-b^2}{a^2+b^2} \rtext{.}
\label{eq:gamma a2b2}
\end{equation}
For $\gamma>1-\kappa$, this becomes a hyperbola.
The position angle of the minor axis is the same as that of the
perturbing mass responsible for the shear ($\theta_\gamma$).
This means, that each ellipse/hyperbola passing through all images
corresponds to 
a value of $\vc\gamma$ with $h/\hnogam=0$. According to equation
\eqref{eq:h crit shear}, for these values
\begin{equation}
\vc\gamma\sprod\vc\gamma\sub c = \gamma\sub c^2
\end{equation}
holds.
The values of $\vc\gamma$ for all these conic sections
span the complete subspace for which $h/\hnogam=0$ (see
Fig.~\ref{fig:gammac}). The 
critical shear is the smallest of these values and can therefore be
calculated with equation \eqref{eq:gamma a2b2} for the `roundest'
ellipse passing through all images.
In Figure~\ref{fig:crit shear}, we show several such ellipses for
some lens systems.

We conclude, that for certain values of $\vc\gamma$
(see~Figure~\ref{fig:gammac}), e.g. for 
$\vc\gamma=\vc\gamma\sub c$, all time delays vanish. 
Each of the fitting ellipses can act
as an Einstein ring for the correct value of $\vc\gamma$. 
The light travel time is the same for
all parts of such rings as Fermat's theorem requires. In our
consideration, the potential is only 
constrained at the image positions and arbitrary in other
directions. The Einstein ring may therefore break up and form a number
of discrete images with still vanishing time delays (see appendix).

\section{Shifting the lensing galaxy}

Surprisingly, most of the quantities we determined do not depend on
the position of the lens centre. A shift of the centre is equivalent
to adding a constant displacement to the vectors $\vc z_i$ and
$\vc\zs$. If we look at the general set of equations \eqref{eq:eqset
  comp}, we see that such a shift adds terms linear in $\vc z_i$ and a
constant to the equations. The constant term can be absorbed in $C$
and (for $\beta\ne 1$) the linear terms in $\vc\zs$. As $C$ and
$\vc\zs$ are of no interest, the equations do not change when applying
this shift.

This means, that $h$ and the critical shear
are translation invariant and can be determined even
if the lens position is not known. This is only true for the family of
models we analyse here. Simple parametric models usually only fit the
data for a specific position of the lens centre.

\section{Spherical models for nearly Einstein ring systems}

For spherical models, the equations are
overdetermined.
There are nevertheless systems, which can be fit accurately with
this kind of model.
It can be shown, that $\hsph/\hnogam$ becomes zero for $\beta\to0$ to
cancel the vanishing denominator in \eqref{eq:H' general new} and
ensure a finite $\hsph$.
It can also be shown, that $\vc\gamma$ is
parallel to $\vc\gamma\sub c$ for arbitrary $\beta$. Taken together, this means
that for point mass 
models, we always have $\vc\gamma=\vc\gamma\sub c$.
The geometrical interpretation can be used to determine the direction
of external shear for spherical models without calculations. It is
given by the orientation of the minor axis of the roundest ellipse
passing through all images. For the systems discussed below (and for
point mass lenses), even the absolute value of $\gamma$ can be determined from
this ellipse.

For systems, where
the $n$ images are all located very close to the Einstein ring at $r_0$,
we can recover another scaling relation.
In this case, the
power-law can be interpreted as a local approximation to {\em any} radial
mass profile, like softened power-law spheres or other models. 
We assume, that one fitting reference model is known. It is then
possible to find 
a family of other models which are also consistent with the
observations.

This was done numerically by \citet{wambsganss94} and led to a
scaling of $\hsph\propto 2-\beta$.
\citet{wucknitz99} presented a simple interpretation of this fact in
terms of the well known mass-sheet degeneracy \citep{gor88}. 
If we multiply the lens equation $\vc\zs = \vc z - \vc\alpha$ with
$1-\kappa$, we get another lens equation with the source position and
lensing potential (or deflection angle) scaled with the same
factor, plus an additional constant convergence $\kappa$. This means,
that lens models given by the deflection angles $\vc\alpha$ and
$\tilde{\vc\alpha}=(1-\kappa)\,\vc\alpha+\kappa\,\vc z$ are equivalent,
but source positions, time delays, etc.\ are scaled by $1-\kappa$ in
the latter model.

\begin{figure}
\begin{center}\includegraphics{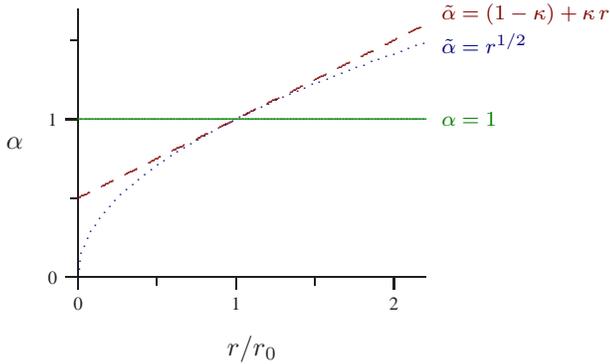}\end{center}
\caption{The deflection angles for the three equivalent lens
  models. The isothermal ($\beta=1$) reference model is shown solid,
  the transformed model dashed and the final equivalent power-law
  model dotted.}
\label{fig:wuref}
\end{figure}%

As reference model, we choose an exponent $\beta_0=1$. This model is
now transformed as just described and then approximated locally near $r_0$
by a modified power law with exponent $\beta$.
Figure~\ref{fig:wuref} illustrates this idea.
For the new exponent, we find
\begin{equation}
2-\beta = 1-\kappa \rtext{.}
\end{equation}
As time delays, source position and external shear all scale with
$1-\kappa$, we find the general scaling laws
\begin{equation}
\vc\zs\propto 2-\beta,\quad \hsph\propto 2-\beta,\quad\vc\gamma \propto 2-\beta
\end{equation}
for an arbitrary mass index $\beta$. The time delay ratios
do not change.

With $\vc\gamma=\vc\gamma\sub c$ for point mass systems, the shear for
arbitrary $\beta$ becomes $\vc\gamma=\vc\gamma\sub c\,(2-\beta)/2$,
leading to a ratio
of the Hubble constants for the spherical and shearless case of
\begin{equation}
\frac{\hsph}{\hnogam} = \frac{\beta}{2}  \rtext{.}
\label{eq:hh0 spec}
\end{equation}
The two models differ significantly for a realistic range of $\beta$.

\section{Influence of the radial mass index $\bmath{\beta}$}

One of the interesting properties of equation \eqref{eq:H'
general new} is the very simple dependence of the results on
$\beta$. The determined value of the Hubble constant $\h$ simply
scales with the factor $(2-\beta)/\beta$.
The most alarming fact is, that this factor does not depend on the
geometry of the lens, the time delay ratios or the amount of external
shear. When using the models described here to determine $H_0$ from
lens time delays, the error due to the assumption of an incorrect $\beta$
will be exactly the same for all lens systems as long as the real $\beta$ is
more or less equal for all lensing galaxies.

\citet{witt00} numerically found a scaling of $(2-\beta)/\beta$ in the case
of a power-law model without external shear for orthogonal image pairs
($\theta_i-\theta_j\approx 90^\circ$), while these computations lead
to $2-\beta$ for opposed images. In contrast to their work, we have used
all time delays as constraints so that they cannot scale
differently. The common scale factor of all time delays shows as a
scaling of $\h$ in our calculations.

The reader might feel as uneasy about the seemingly diverging time delays or
$\h$ in the limit $\beta\to0$ as the author did.
This limit is equivalent to point mass models and one should not observe
diverging time delays for this kind of lens. The
point causing trouble here is the fixed external shear in our
considerations. To fit the data with a point mass model, the shear has
to be equal to the critical shear $\vc\gamma\sub c$. Taking this into
account, the result gets multiplied by a vanishing
$\h/\hnogam\propto\beta$ which cancels the $1/\beta$ factor.
If we now change the shear by a small amount, the potential and thus
$\h$ will immediately diverge in the limit of small $\beta$. This has
no direct physical implications, because the mass models 
will become extremely unphysical to compensate for the shear
effect. Small relative differences in the $F_i$, that will be
introduced by an incorrect shear, will lead to enormous asymmetries in
the mass distribution. This is related to the fact, that realistic compact mass
distributions can provide only almost spherically symmetric
potentials. Any multipole moments would radially decrease more rapidly than the
monopole term and must be very strong to have any effect.
We will discuss this in more detail for a special case in
section~\ref{sec:einstein symmetry}.

For spherical systems with images near the Einstein ring, we confirmed
the approximate 
scaling of $\hsph\propto 2-\beta$. This seems to be incompatible with the
general scaling law of $\hnogam\propto(2-\beta)/\beta$ at first sight.
With the factor of $\hsph/\hnogam=\beta/2$ from equation \eqref{eq:hh0
  spec}, by which the shear changes the result in spherical models,
these two results are, however, in perfect agreement.

\section{Application to special cases}
\label{sec:special cases}

To illustrate the results, we have presented here, and to test their
relevance for real lenses, we want to
apply the formalism to systems with a special Einstein cross like
symmetry and to some 
of the known real systems that are either useful to actually determine
$H_0$ (1115+080, 0911+0551, 1608+656) or are interesting because
they are very well studied systems like 2237+0305. The calculations
will show, that both scaling relations ($\hsph\propto 2-\beta$ and
$\hnogam\propto (2-\beta)/\beta$) are relevant for the determination
of $H_0$ from time delays. The detailed numerical models for
2237+0305 will furthermore show, that the scaling also applies if
the time delays themselves are not used to constrain the models. All
time delays scale almost exactly as predicted by our analytical
work. We will also see that parametric models may fit only for a
limited range of $\beta$. The scaling relations are then valid only
within this range.

Besides the effects of the exponent $\beta$, the possible strong
effects of any external shear will also be confirmed by the numerical models.

\subsection{Symmetric Einstein cross like systems}
\label{sec:einstein symmetry}

A rather special example of systems shall be discussed explicitly in this
section. We consider a lens with time delays $\Delta t_{12}=\Delta
t_{34}=0$ and the following image positions:
\begin{alignat}{2}
x_{1,2} &= \pm(1-\epsilon)\,r_0  \qquad & y_{1,2} &= 0 \\
x_{3,4} &= 0                            & y_{3,4} &= \pm
(1+\epsilon)\,r_0 
\end{alignat}
From these data, we
immediately conclude $\vc\zs=0$. The $y$-component of the shear does
not contribute at all and cannot be constrained. The equation
determining the Hubble constant now reads
\begin{equation}
\h\,\Delta t_{13} = (1-\kappa)\,r_0^2\,\frac{2-\beta}{\beta}
  2\,\epsilon \left(1-\frac{\gamma_x}{\gamma\sub c} \right) 
  \rtext{,}
\label{eq:einstein gen}
\end{equation}
with the critical shear
\begin{equation}
\gamma\sub{c} = (1-\kappa)\,\frac{2\,\epsilon}{1+\epsilon^2} \rtext{.}
\end{equation}
The special symmetry makes it possible to choose any value of external
shear even for isothermal models.
Furthermore, it is possible to exactly reproduce the data with
spherically symmetric 
models plus external shear. In this case,
the shear is uniquely defined:
\begin{align}
\gamma 
&= (1-\kappa)\,(2-\beta)\,\epsilon + {\cal O}(\epsilon^2)
\label{eq:gamma scale spherical}
\end{align}
The time delay equation now becomes somewhat more complicated than in
the non-spherical case. To first order in $\epsilon$, it reads
\begin{align}
\hsph\,\Delta t_{13} 
 &= (1-\kappa)\,r_0^2\,(2-\beta)\,\epsilon + {\cal O}(\epsilon^2) 
\rtext{.}
\end{align}
Comparing this with \eqref{eq:einstein gen}, we
recover the factor of $\beta/2$ between the spherical and
shearless case.
Both models fit the data exactly and, in
this special case, the models are even compatible with highly popular
elliptical mass distributions. That means that without any independent
information about the external shear (or equivalently the ellipticity
of the galaxy itself), we have a factor of two uncertainty even when
only considering these two simplest models for $\beta=1$. The real
situation may 
be even much worse, when we consider models with internal {\em and}
external shear. In this case, any small unknown contribution of
external shear of the order of $\gamma_c$ (which for very symmetric
systems becomes arbitrarily small) will change the result significantly.

\citet{witt95} discussed exactly the same type of systems
with spherical models plus shear. With $\gamma$ fixed, they also
derived a scaling law of $(2-\beta)/\beta$ (see their equation
8\footnote{The exponent of the first term in equation (8)
in \citet{witt95} is incorrect, it should be the same as that
in the second term. (S.~Mao, private communication)}). When the shear is constrained by the lens
equations, the scaling changes to the $2-\beta$ form.

We finally want to discuss the consequences of diverging time delays
in the fixed shear 
case for $\beta\to0$  due to equations \eqref{eq:H' general new} and
\eqref{eq:einstein gen}.
For 
simplicity, we 
assume $\vc\gamma=0$, $\kappa=0$ and $\epsilon\ll1$, but the argument
is generally also true for other values. 
We write the potential as a multipole expansion\footnote{We use an
  expansion for the 
  principal axes of the lens system. The $\sin2k\theta$ terms might be
  included as well but they would not change the density on these
  axes, which is what we are interested in.}
\begin{equation}
\beta\,F (\theta) = r_0^{2-\beta} \left(1-\sum_{k=1}^\infty a_k
\cos 2k\theta\right) \rtext{.}
\end{equation}
To be compatible with equation \eqref{eq:F}, the coefficients
have to meet the condition
\begin{equation}
\sum_{k=1}^\infty a_k = (2-\beta)\,\epsilon\rtext{.}
\end{equation}
We notice, that densities \eqref{eq:sigma} can become negative near the
axes. To minimize the angular density contrast, we have to
keep only the monopole and quadrupole terms and set all higher
coefficients to 0.
The potential is then equivalent to a density of
\begin{equation}
\sigma = \frac{\beta}{2}\Bigl(\frac{r}{r_0}\Bigr)^{\beta-2} \biggl(
  1+ \epsilon\,(2-\beta)\,\Bigl(\frac{4}{\beta^2}-1\Bigr)\cos2\theta
  \biggr)
\rtext{,}
\end{equation}
which is everywhere positive only for sufficiently high values of
$\beta$. When using realistic mass models, we can
therefore expect a lower bound for $\beta$ to achieve acceptable fits.
This applies not only to this special symmetric lens system but is
true in general.
Numerical models presented in the next section will confirm this
result for $\gamma=0$ (see Figure~\ref{fig:timdel 2237}).

\subsection{The Einstein cross Q2237+0305}
\label{sec:2237}

This lens is not usually taken into consideration when thinking about
determination of $H_0$, because the time delays are expected to be
very small and can therefore not be determined easily.
Here we show that even if all three time delays were known
exactly, constraints for the Hubble constant would still be very weak.

The degeneracy caused by the unknown mass index $\beta$ was already
discussed by \citet{wambsganss94} for spherical models plus external
shear. The authors found the $2-\beta$ scaling using numerical models.
We now want to investigate how strong the assumption of a spherical
main galaxy really influences the results.

No time delays are available for 2237+0305 and they may never
be determined. We can nevertheless calculate the critical shear defined
before and compare it with the a typical value one gets for spherical
models.
Positions including error bars used for this were taken from
\citet{crane91} to make results comparable with \citet{wambsganss94}.

The critical shear as calculated from these positions\footnote{All
  coordinates in this paper: $x$ to east and $y$ to north.} is 
$\vc\gamma\sub c = (0.0915\pm0.044, 0.0958\pm0.045)$ or
$|\vc\gamma\sub c| = 0.13 
$ where the errors are $1\sigma$ bounds from Monte Carlo
simulations. Numerical modelling
results in a shear of $0.0696$ almost exactly parallel to
$\vc\gamma\sub c$  for isothermal spherically symmetric
potentials. We therefore expect the time delays (or $\h$ if we take
$\Delta t$ as known) of the spherical model with shear to be a factor
$0.4745$ smaller than in the shearless case. For the moderately small
shear of $\approx 0.07$, this is a huge effect. This factor is in good
agreement with the expected value of $1/2$ for idealized systems.

To compare results in the general case, we performed
numerical model fitting with an elliptical potential approach plus
external shear.
\begin{align}
F(\theta) &\propto \left( \Bigl(\frac{\cos\theta}{1+\epsilon}\Bigr)^2 +
\Bigl(\frac{\sin\theta}{1-\epsilon}\Bigr)^2 \right)^{\frac{\beta}{2}}
\end{align}
Elliptical potentials are known to be unphysical for large
ellipticities. Although $\epsilon$ is small in our case, we may expect
unrealistic solutions for small values 
of $\beta$, because the limit of acceptable ellipticities vanishes for
$\beta\to0$ (cf. last section). In fact the fitted $\epsilon$ also
increases with decreasing $\beta$.

\begin{figure}
\includegraphics[width=8.2cm]{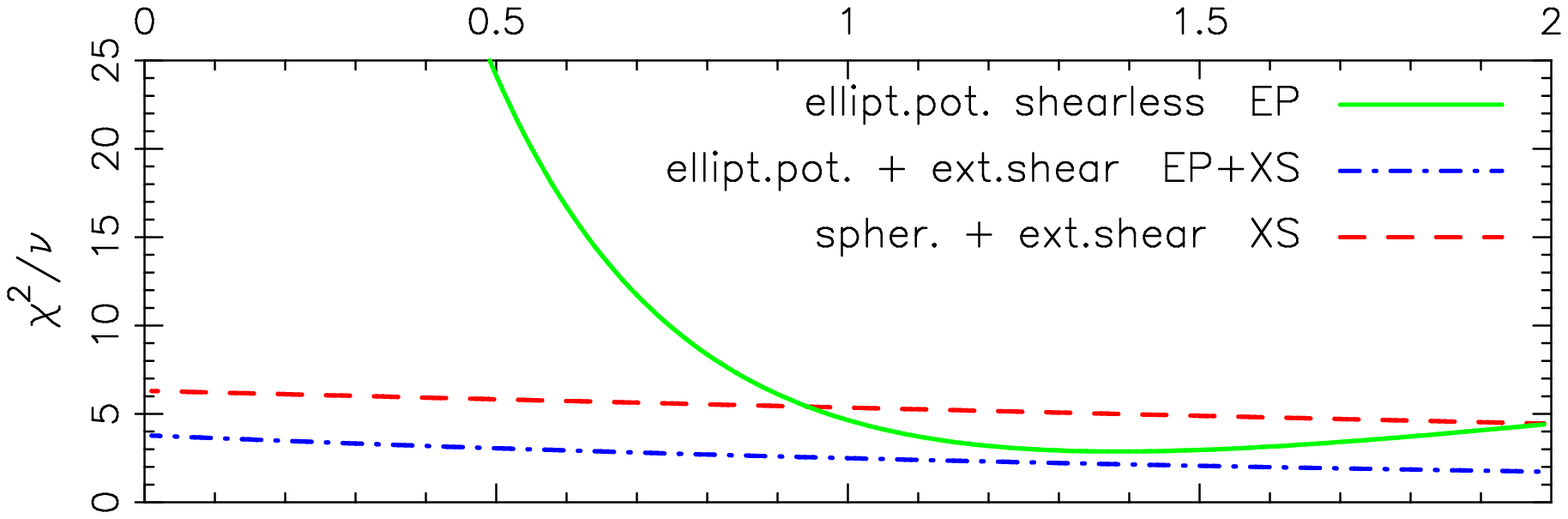}\\
\includegraphics[width=8.2cm]{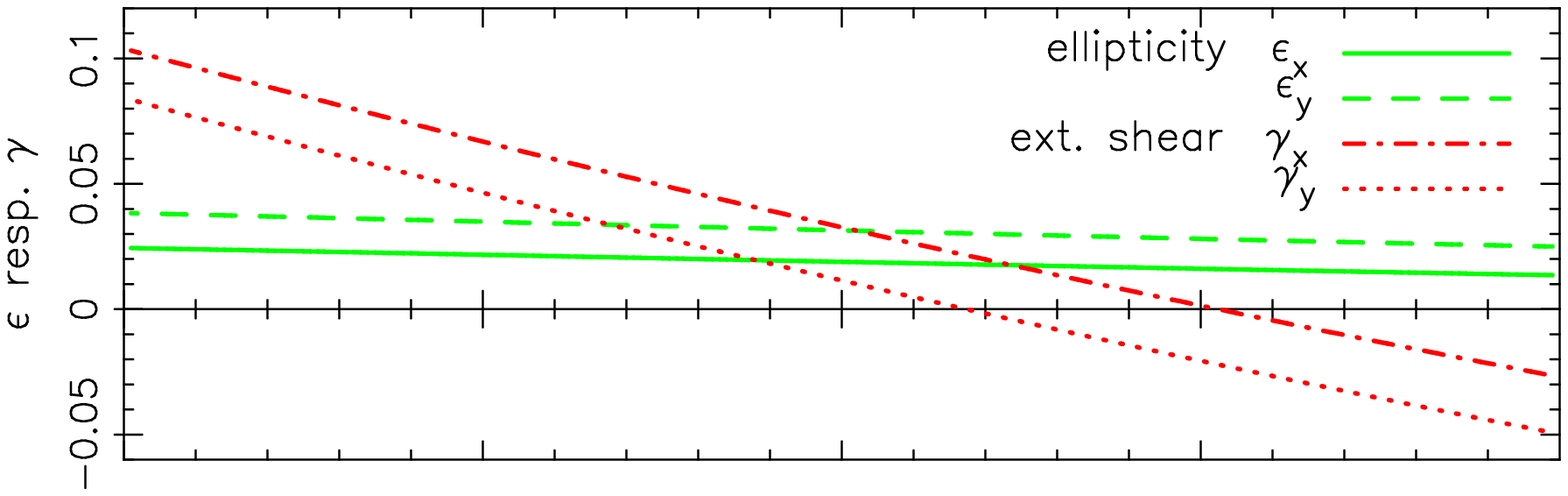}\\
\includegraphics[width=8.2cm]{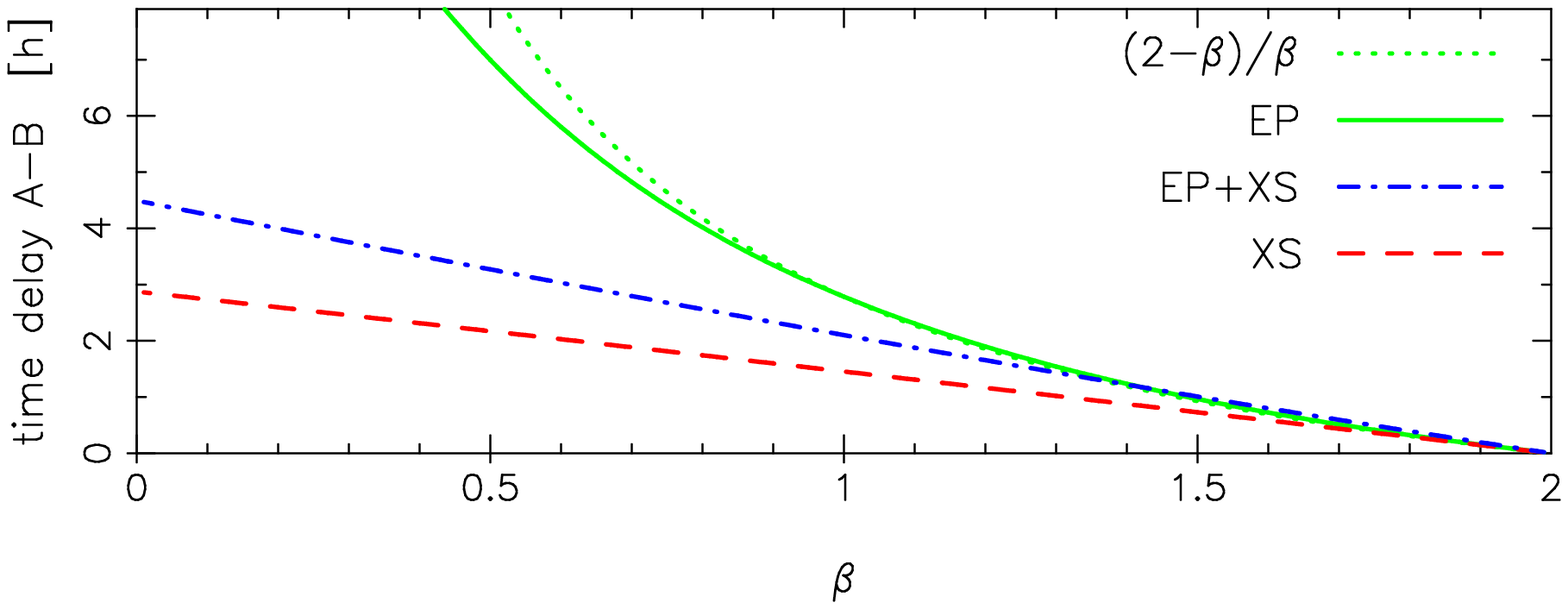}
\caption{Results of numerical model fitting. The top panel shows the
  reduced $\chi^2$ for the best models. Here $\beta$ was
  counted as fixed, resulting in $\nu=3\,/\,3\,/\,1$ degrees of freedom for the
  spherical\,/\,shearless elliptical potential\,/\,general elliptical
  potential model, denoted as XS\,/\,EP\,/\,EP+XS. In 
  the middle, both components of the shear and ellipticity are
  plotted for the EP+XS model. The last panel presents
  the time delay A$-$B in hours for
  $H_0=75\,\kmsmpc$ and Einstein-de~Sitter
  cosmology. We also included a curve scaling with $(2-\beta)/\beta$
  to compare with the EP case.}
\label{fig:timdel 2237}
\end{figure}

We decided not to use invented time delays (calculated for a reference
model) to fit the models.
In this way, we can check the validity of our results even for cases
where multiple time delays are not used as constraints.
Plots of the residuals, ellipticity, shear and time delay between component
A and B are shown in 
Figure~\ref{fig:timdel 2237}.

The non-vanishing residuals at $\beta\to2$ might be worrying at first,
because a spherical model without external shear can fit any image
configuration for $\beta=2$. The result would be a sheet of
constant density equal to the critical density. There would be no
isolated images but an area of constant (and very high) surface
brightness. One might thus naively think, that residuals should be very
small near $\beta=2$. This is not the case. It is true, that
the deviations of the projected images will become arbitrary small in
the source plane. On the other hand,
however, the magnifications diverge in the limit, causing the
deviations in the lens plane to stay finite.

Three families of models have to be discussed.
First, we fixed the ellipticities at 0 to compare with the
results from \citet{wambsganss94}. The residuals are almost constant
for spherical models. The shear and the time delays scale
very accurately with $2-\beta$ as in the idealized
considerations.

More interesting in our context is the behaviour in the shearless case
where we expect to find a $(2-\beta)/\beta$ scaling of the time
delays. The $\chi^2$ gets unacceptably large for $\beta\ll 1$. In the
other cases ($\beta\ga 0.6$), the agreement with the theoretical
predictions also shown in Fig.~\ref{fig:timdel 2237}
is very good. For isothermal models, the ratio of time delays
calculated for the two models (spherical$/$shearless) is 10--20~per~cent larger than predicted
by the critical shear. This is still a very good agreement
considering that the time delays were not used as constraints for the
numerical models.

Real lensing galaxies usually are elliptical and also embedded
in an external shear field. We therefore also let both $\epsilon$ and
$\gamma$ vary freely in order to minimize $\chi^2$. Counting the formal
number of constraints and parameters, we expect a minimum of
$\chi^2=0$. The fact, that $\chi^2$ does not vanish is a confirmation
of the degeneracy involving shear and ellipticity already discussed by
\citet{witt97}. The effective number of parameters is therefore
smaller then the formal one.
Because we notice the ellipticity changing only slightly with $\beta$,
a scaling like that in the spherical case is expected. This can indeed
be seen in the plot, where the time delay scales with $2-\beta$. Contrary
to the spherical case, $\vc\gamma$ does not scale proportional to
$2-\beta$ but is additionally shifted by a constant offset. This is due to the
fact that part of the shear has been transformed to ellipticity.

We conclude, that both scaling laws for $\h$ or the time delays can be
relevant, depending on the family of models used. In the case of
2237+0305, however, the influence of external shear is stronger
than the effect of $\beta$ for any realistic values of the latter.

\subsection{PG~1115+080}

Models and their degeneracies for this quadruple system have been studied
extensively
\citep[e.g.][]{courbin97,impey98,keeton97,saha97,schechter97,zhao01}, 
leading to a variety of more or less realistic mass distributions and
a range of values for the Hubble constant between about 40 and
$80\,\kmsmpc$.

All these authors agree on the importance of taking the effect of a
nearby galaxy group into account. Our formalism includes this group as an external shear of the
order $\gamma\approx 0.1$. This shear is however not well
constrained. \citet{keeton97sel} showed, that the residuals do not
change much for ranges of $\gamma\approx0.06$--$0.2$. With our
general family of models, the effect of unknown external shear can be
quantified by using the critical shear and equation \eqref{eq:h crit
  shear}. To calculate $\gamma\sub c$, we only need image positions
relative to an arbitrary reference centre. The uncertainty in the
galaxy position, which is usually much higher than in the image
positions, does not affect the result.
Using the HST observations from \citet{impey98} with their claimed
accuracy of $0\farcs002$ as a basis for Monte Carlo simulations, we obtain
a critical shear of $\vc\gamma\sub
c=\twovectext{0.142\pm0.002}{0.167\pm 0.003}$. Although the ground
based positions from 
\citet{courbin97} are not compatible with the HST results within the
formal error bars, the critical shear from these data is about the
same, $\vc\gamma\sub c=\twovectext{0.143\pm0.003}{0.160\pm
  0.005}$. This means, that any shear of the order 0.1 can 
change the results for $H_0$ significantly. As the uncertainties in
$\gamma$ are of this order of magnitude, large effects on $\h$  result
even for fixed $\beta$.

The formalism we present in this article was not developed to directly
determine $H_0$ from observations but to study the model degeneracies
and scaling laws. One might nevertheless try to use the prescription
from section~\ref{sec:iso} to obtain an estimate for the
Hubble constant and the 
external shear for an isothermal model. The
errors of the observational data of course have to be taken
into account.

Time delays derived from the same light curves have been published by
\citet{schechter97} and \citet{barkana97}. The A component was not
resolved in these observations, and time delays were only determined
relative to the sum of A$_1$ and A$_2$. This is justified by the
small time delay between the two, which is expected to be the order of hours.
For our Monte Carlo simulations, we assumed a time delay between the A
images of $(0\pm1)$\,days. 

Results for $H_0$ in the isothermal case differ depending on which set
of time delays and positions is 
used. With the redshifts $z\sub s=1.722$ and $z\sub d=0.310$
\citep{tonry98} and an 
Einstein-de~Sitter universe, we obtain values of $H_0$ between 47 and 
$58\,\kmsmpc$ with errors between 12 and
30~per~cent ($1\sigma$).  The external shear is only weakly
constrained, but seems to be of the order 0.1.
We have to stress, that this result includes
all possible isothermal models with arbitrary angular dependence and
is thus much more general than elliptical mass distributions. As long
as different determinations of the positions and time delays are not consistent
with each other within their error bars, any results for $H_0$ have to
be interpreted with care of course.

\subsection{RX~J0911.4+0551}

This quad, initially discovered as a triple \citep{bade97}, has unique
geometrical properties and is a strong candidate for time delay
determination, although no result has been published yet. Rapid
variability has been detected in the X-ray regime,
providing the possibility of a determination of all three time delays with
unprecedented accuracy \citep{chartas01}.
Our first
models (elliptical potential plus shear, $\chi^2\approx1$) were presented in
\citet{burud98}. The external shear in the best-fitting model is
$\gamma=0.32$ and points almost exactly in the direction of a nearby
cluster of galaxies.

The redshift and velocity dispersion of this
cluster was measured by \citet{kneib00} to $z=0.769$ and
$\sigma_v=(836\pam{180}{200})\,\rmn{km\,s^{-1}}$. From this, they
derive an absolute shear of $\tilde\gamma=0.11\pam{0.05}{0.04}$ using
a SIS model for the cluster.

To compare with lens models, we have to use the reduced shear
$\gamma=\tilde\gamma/(1-\kappa)$, because the convergence $\kappa$ caused by
the cluster was not taken into account explicitly. The mass-sheet degeneracy
simply scales all parameters in \eqref{eq:eqset comp} with $1/(1-\kappa)$.
For SIS models,
$\tilde\gamma=\kappa$ holds and we obtain $\gamma=0.12\pam{0.05}{0.04}$.
This measurement is not
in good agreement with the model from \citet{burud98}. The reduced shear 
differs by about 0.2, the direction of the two being almost
identical. A possible explanation for this discrepancy is the presence
of a second galaxy close to the main lens that might change the
potential considerably. It is also possible that the internal
asymmetry of the main galaxy itself can not be described as an
elliptical mass distribution.

To estimate the uncertainty in time delays or the Hubble constant
derived from them, we use again the critical shear which is
$\vc\gamma\sub c=\twovectext{-0.553\pm0.013}{0.101\pm 0.005}$ or
$|\vc\gamma|=0.56$. Even with this very large critical shear, the
uncertainty in the real $\gamma$ has significant effects, because it
is large as well.

\subsection{B1608+656}

This system is the first and up to now only quad for which all three
independent time delays have been measured \citep{koopmans99i}. This
offers the unique possibility to apply our method to a system
providing the complete set of constraints. HST images
show a main lensing galaxy but also a weaker
second galaxy between the four images \citep{jackson98}. We
nevertheless apply the 
method to B1608+656, if only to see if the effect of a secondary lens
can be detected in this way, e.g.\ by pretending there is a very large external
shear.
Data for the positions of the images and the main lensing galaxy were
taken from \citet{koopmans99ii}. The formal accuracy of the image positions is
extremely high, of the order 2--12\,$\umu$arcsec. For the Monte
Carlo simulations, we used 1\,mas scatter in each coordinate to account for
possible shifts by local density fluctuations, caused for example by
globular clusters \citep{mao98}.

We can use the general equation \eqref{eq:eqset comp} to determine the
Hubble constant for shearless models. With the redshifts of $z\sub
d=0.6304$ and $z\sub s=1.394$ and standard Einstein-de~Sitter
cosmology, we obtain a value of $H_0=
(37\pm5)\,\kmsmpc$ for $\beta=1$. The critical shear is $\vc\gamma\sub c =
\twovec{0.072\pm 0.001}{0.069\pm0.001}$.
For isothermal models, equation \eqref{eq:eqset comp} predicts a shear of
about
$\vc\gamma = \twovec{-0.32\pm0.02}{-0.11\pm0.01}$, depending on which
HST image is used to determine the galaxy position. The result for the
Hubble constant is $H_0=(130\pm15)\,\kmsmpc$.

The enormous differences in both $H_0$ predictions is a consequence of
the large external shear $\gamma\gg\gamma\sub c$. No external shear at
all is
needed to fit the data when including the influence of the second
galaxy in the field.  The models in \citet{koopmans99ii} even predict
velocity dispersions for both galaxies which are of the
same order of magnitude. This is surprising, since the secondary galaxy
is much weaker in all bands in the optical images.

\section{Discussion}

We used a very general semi-parametric lens model approach to study
the changes of time delays and the determined Hubble constant with the
assumed radial density slope, quantified by the radial mass index
$\beta$. By using only linear constraints, it was possible to keep
all the fundamental equations linear. This made the study of
the $\beta$ dependence easy. For fixed external shear in quadruple lenses, this resulted
in the simple scaling law $H_0\propto (2-\beta)/\beta$, independent
of the lens geometry, the time delay ratios or the external shear. This
means, that a systematic error in the assumed $\beta$ will have exactly
the same effect on all lenses and will not show as scatter in the results.
The good agreement between $H_0$ measurements from different lenses
\citep{koopmans99ii} should therefore not be 
taken as evidence for an accurate determination of $H_0$. It merely
shows, that all lensing galaxies seem to have more or less the same
$\beta$.

In nearly
isothermal models, a systematic error of only 10~per~cent in $\beta$ will
result in an error of about 20~per~cent in the deduced Hubble constant. To
compare the results from lensing not only with each other but also
with results derived from other methods, this possible source of
error has to be taken into account.

Furthermore, it is important not only to be aware of this effect, but to
try and obtain better constraints on the radial mass profile.
Possible ways to do 
this include modelling lenses with multiply imaged
extended sources to obtain constraints for the lensing potential at
a wide range of distances to the lens centre. Lenses with
multiply imaged point sources can constrain the models only at a small
number of radii.
This
should be done not only for systems with measured time delays, but for
as many applicable lenses as possible to acquire reliable {\em
measurements} of $\beta$ for a representative sample of lenses.

Another possibility is the detailed study of the dynamics of
lensing galaxies or of other (usually low redshift) galaxies to learn more
about the range of realistic galaxy mass profiles and use the results
of this analysis to produce more realistic models for the lenses.

We also
quantified the effect of external shear by introducing the concept of
a `critical shear' $\vc\gamma\sub c$. The effect of $\gamma$
on $H_0$ is linear and strongest in the direction of $\vc\gamma\sub
c$. For a fixed direction, its amount is proportional to $\gamma/\gamma\sub c$.
The shear has to be exactly critical to fit the
observations in point mass lenses. 
The value of $\vc\gamma\sub c$ can be found in a geometrical way. It
is given by the ellipticity of the roundest ellipse passing through
all images. For $\vc\gamma=\vc\gamma\sub c$ and non-vanishing $\beta$,
the time delays $H_0\Delta t$ become
zero. This is also true for a whole family of models which are
represented by the less symmetric ellipses fitting the images.
In the appendix, we discuss the relation to possible Einstein rings
which are exactly given by these ellipses.

The effect of shear is also the clue in understanding the
compatibility of the general scaling law with the simpler one of
$\hsph\propto2-\beta$ for spherical models. In the latter models, the
shear is constrained by the observational data and changes $\h$ by a
factor of $\beta/2$ when compared to shearless models.
In cases where spherical models are able to
fit the data, the allowed range of $\gamma$ results in an uncertainty
of $\h$ always covering both models.
This may in certain cases only apply for a limited range of $\beta$.

Interestingly, the value of the critical shear and the Hubble constant
in the general model (with fixed shear for $H_0$) do not depend on the
position of the lensing galaxy. This may be of use for systems where
this position cannot be determined accurately.

When using more general models or less constraints than in our
calculations, the scaling laws still apply. They are only valid for a
subset of the possible models then and one would have to expect even
larger uncertainties when using the whole set. This also applies for
lenses with less than four images.
More special models on
the other hand, like parametrized elliptical power-law models, may be
able to constrain the range of possible results much
better. Nevertheless, the scaling laws still apply for the range of
models that are compatible with the constraints.
Even in these cases, our results may be used to determine $H_0$ without
explicit modelling.

\section*{Acknowledgments}

It is a pleasure to thank Ester Piedipalumbo, Shude Mao and Sjur
Refsdal for interesting discussions on the subject.
The very constructive comments from the referee P.~Saha greatly helped
in understanding the geometrical properties of the critical shear.



\appendix

\section{Einstein rings and high image multiplicities}

An interesting property of the general power-law models we used in the
main part of this paper is the possibility to produce Einstein rings
from point sources for arbitrary values of the external shear.
For an Einstein ring parametrized by $r=r(\theta)$, all points on this
ring must have the same light travel time to meet Fermat's theorem.
For this appendix, we set $\kappa$ and $\gamma_y$ to 0  for
simplicity. For an arbitrary direction of the shear, we just
have to replace $\theta$ by $\theta-\theta_\gamma$. We also assume,
that $\gamma<1$.

With $t=0$, the general equations \eqref{eq:eqset comp} describe an
ellipse, which is centred on the lens in the special case $\vc\zs=0$
or $\beta=1$.
\begin{equation}
r = r_0\,(1+\gamma\cos2\theta)^{-1/2}
\end{equation}
The minor and major axes are $r_0/\sqrt{1\pm\gamma}$, compatible with 
equation \eqref{eq:gamma a2b2}. The potential is in this case an elliptical one
\begin{equation}
F (\theta) = \frac{1}{\beta}
\,r_0^{2-\beta}\,(1+\gamma\,\cos2\theta)^{\beta/2} \rtext{.} 
\label{eq:global ring F}
\end{equation}
With respect to the tangential caustic, the effects of ellipticity and
shear cancel in these models and the caustic degenerates to a point.
This is qualitatively different from elliptical mass distributions
where the caustic is deformed and overlaps itself, producing areas of
higher multiplicities (see below).
With arbitrary
$\vc\zs$ and $\beta$, the centre of the ellipse is shifted to $\vc
z_0$ with\footnote{This equation is valid for arbitrary directions of $\vc\gamma$.}
\begin{equation}
\matr{1+\gamma_x & \gamma_y \\ \gamma_y & 1-\gamma_x} \vc z_0 =
\frac{1-\beta}{2-\beta} \vc\zs \rtext{.}
\end{equation}
For $\beta\ne1$, this shift can take any value if $\vc\zs$ is varied.
To obtain a globally unique function $F(\theta)$, the centre of the lens
has to be located inside of the ellipse.

Even for lens systems with four images, it is always possible to find an
ellipse passing through all of them, which can act as an Einstein ring for the
corresponding value of $\vc\gamma$ given by the ellipticity.
This does not mean, that we always see an Einstein ring for this
special value of external shear, as $F(\theta)$ is not constrained for
angles between the images.

Small deviations from the Einstein ring case can lead to an arbitrary
number of images near the former elliptical ring. Special cases of
these systems (singular isothermal ellipsoidal mass distributions with
shear) with up to eight images were mentioned by
\citet{lopez98} and \citet{witt00b} and discussed in detail by
\citet{keeton00}.
\citet{evans01} present results for shearless models with arbitrary
$F(\theta)$.
From \eqref{eq:F} and \eqref{eq:F'} we obtain the following condition
for an image of a source at $\vc\zs=0$:
\begin{equation}
\frac{F'}{F} =
-\frac{\beta\,\gamma\,\sin2\theta}{1+\gamma\,\cos2\theta}
\label{eq:diff eq images}
\end{equation}
A global solution for this differential equation is given by
\eqref{eq:global ring F} which leads to the elliptical ring we
discussed before. 
For a number of discrete images, a more general solution is possible:
\begin{equation}
F (\theta) = f(\theta) \,(1+\gamma\,\cos2\theta)^{\beta/2}
\end{equation}
At the positions of the images $\theta_i$, \eqref{eq:diff eq images} has to be
met, leading to the simple condition
\begin{equation}
f'(\theta_i) = 0 \rtext{.}
\end{equation}
As $f(\theta)$ is an arbitrary function, we can easily construct systems with
any number of images.
The radial coordinates of the images can then be determined to be
\begin{equation}
r (\theta_i) = \bigl(\beta \,f(\theta_i)\bigr)^{1/(2-\beta)}
(1+\gamma\,\cos2\theta_i)^{-1/2} \rtext{.}
\end{equation}

\bsp

\end{document}